\documentstyle[aps,preprint,pra]{revtex}
\tighten
\renewcommand{\epsilon}{\varepsilon}
\begin{document}
\newtheorem{thm}{Theorem}
\newtheorem{prop}[thm]{Proposition}
\newtheorem{cor}[thm]{Corollary}
\newtheorem{lemma}[thm]{Lemma}

\draft
\title{Method for decoupling error correction from privacy amplification}
\author{Hoi-Kwong Lo\footnote{email: hoi\_kwong@magiqtech.com}}
\address{MagiQ Technologies, Inc., 275 Seventh Avenue,
26th Floor, New York, NY 10001-6708}

\date{\today}
\maketitle
\begin{abstract}
Entanglement purification provides a unifying framework for
proving the security of quantum key distribution schemes.
Nonetheless, up till now, a {\it local} commutability
constraint in the CSS code construction means that the error
correction and privacy amplification procedures of BB84 are not fully
decoupled. Here, I provide a method to decouple the two processes
completely. The method requires Alice and Bob to share
some initial secret string and use it to encrypt the
bit-flip error syndrome using one-time-pad encryption.
As an application, I prove the
unconditional security of the interactive Cascade protocol,
proposed by Brassard and Salvail for
error correction, modified by one-time-pad encryption of the error syndrome,
and followed by the random matrix protocol for privacy amplification.
This is an efficient protocol in terms of both computational
power and key generation rate.

\end{abstract}

\smallskip
\pacs{\noindent
 \begin{minipage}[t]{5in}
 \end{minipage}
}

\section{Introduction}
\label{S:Intro}

An important application of quantum information processing is
quantum key distribution (QKD) \cite{BB84,ekert}. The goal of
QKD is to allow two
communicating parties to detect any eavesdropper. Unlike conventional
key distribution scheme, QKD makes no assumptions on the
eavesdropper's computing power. Rather, the security of QKD is
supposed to be based on the fundamental laws of quantum mechanics.

Security proofs of QKD is an important but difficult problem in quantum
information theory. Recently, entanglement purification
\cite{BDSW,BBPSSW} has
become a fruitful avenue of studying the security of QKD.
Roughly speaking, entanglement purification is a generalized form of
quantum error correction for a quantum {\it communication} channel,
rather than quantum storage which is dealt with by
standard quantum error correction.
It was first suggested by Deutsch {\it et al.} \cite{deutsch}
that entanglement purification procotols (EPPs) can correct errors
introduced by the eavesdroppers and allow the two communicating
parties, Alice and Bob, to obtain perfectly entangled (i.e.,
quantum-mechanically correlated) quantum systems, so-called EPR pairs,
from which they can generate a secure key.

A proof of security by Mayers \cite{mayersqkd} applies to a
standard QKD scheme, BB84 \cite{BB84}, published by
Bennett and Brassard in 1984. 
Mayers' proof makes no explicit reference to entanglement purification,
but is rather complex.
A proof of security of QKD based on entanglement purification
has been provided by
by Lo and Chau \cite{qkd}. It has the advantage of being intuitive and
conceptually simple, but it requires that
Alice and Bob possess quantum computers for its implementation.
Recently, Shor and Preskill\cite{shorpre} has removed this
requirement and applied
the approach of entanglemen purification to prove the
security of BB84 \cite{BB84}.
Other proofs of security of QKD that make no explicit
reference to entanglement purification include \cite{biham,benor}.
Recently, a security proof with a practical set-up (weak coherent
states, lossy channels and inefficient detectors, etc) has been
presented by Inamori, L\"{u}tkenhaus and Mayers \cite{ilm}.

Recall that error correction and privacy amplification are necessary
in the generation of the final secure key from the raw quantum
transmission date. Error correction ensures that Alice and
Bob will share a common string and, roughly speaking,
privacy amplification ensures that Eve most likely knows
almost nothing about the key.
Unfortunately, so far the application of entanglement purification
approach to QKD implies a non-trivial constraint between the
two processes, namely the corresponding measurement operators
employed by Alice and Bob must be {\it locally commuting}.
Such a local commutability constraint means that the two
processes are not totally decoupled from each other.
Therefore, it is not entirely obvious how to study error correction
and privacy amplification independently.

In this paper, I propose a novel method to remove this
local commutability constraint, thus allowing us to
decouple the error correction process from the
privacy amplification process. This amounts
to much simplification in the study of both processes.
In the EPP picture, the proposed method requires Alice and Bob to share some
ancillary pre-distributed pure EPR pairs. Instead of measuring the
(bit-flip) error syndrome directly, each user collects the
output into those ancillary EPR pairs and measures those pairs.
(A specific instance, so-called breeding method, of
such a general method, was used in \cite{BDSW}.)
In the BB84 picture, the proposed method requires Alice and Bob to
share initially some common secret ancillary binary string, $a$.
Instead of annoucing the bit-flip error syndrome, which is a binary string
$x$, each user encrypts the error syndrome bit-wise using $a$ as a
one-time pad and announces the encrypted version,
$y = x + a~(mod~2)$, bit-wise.

As an application of the proposed method, I consider
a rather general class of classical error correction methods---the
so-called {\it symmetric} stabilizer-based\footnote{By
stabilizer-based, I only mean that each operator that Alice
measures is a Pauli operator. The various operators are {\it not}
required to commute.} schemes---which may involve
either one-way or two-way classical communications. I show that
any symmetric stabilizer-based scheme
can be modified and subsequently combined with
{\it any} {\it symmetric} stabilizer-based
privacy amplification procedure into an unconditionally secure
protocol for QKD. This means that one can study the
two processes---error correction and privacy amplification---independently.
Such a decoupling of error correction from
privacy amplification allows one to simplify the analysis of security of
a general error correction scheme.

As an application,
I prove the unconditional security of a modified version of the
Cascade scheme \cite{cascade} for error correction
invented by Brassard and Salvail, (followed by, for example,
a random hashing procedure for privacy amplification\cite{mayersqkd}).
This is the first time such a
computationally efficient scheme has been proven to
be secure. Therefore, the result is of practical interest.

Finally, note that the proposed method can be employed as a
sub-routine in concatenated entanglement purification procedures,
including those involving two-way classical communications, as
studied by \cite{twoway} and those
involving degenerate
codes \cite{sixstate}.

\section{Motivation}
\label{s:motivation}
A key motivation of this work is to provide a rigorous
proof of security of interactive protocols for
error correction in QKD. Let me explain in detail.
In QKD, one often has to perform error correction at
a rather high bit error rate of say a few percents, which is much higher
than the typical value of say $10^{-5}$ in conventional communications.
Moreover, one would like the key generation rate to remain high.
As a rule of thumb, the fewer bits are exchanged
between Alice and Bob, the higher the key generation rate.
Furthermore, one would like to implement a QKD scheme efficiently.
That is to say with a minimal amount of computational power.
In a general implementation of QKD,
it is a highly complex question what the trade-off between
the various parameters---tolerable error rate, key generation
rate, computational power---would be the best.

Forward error correction is commonly employed in
conventional communications and works efficiently at
low error rates.
Unfortunately, QKD has a high bit error rate. If forward
error correction is employed in QKD,
a very large block size of order $10^5$ would probably be needed.
This translates to a large amount of computing power.\footnote{I thank
enlightening discussions with Tsz-Mei Ko and Norbert L\"{u}tkenhaus
on this point.}

Two-way communications between Alice and Bob are useful in reducing the
required computing power for error correction. In the
literature, several interactive protocols
such as ``BBBSS'' \cite{BBBSS} and ``Cascade'' \cite{cascade}
have been proposed for error correction
in QKD.\footnote{I thank Norbert L\"{u}tkenhaus for providing the
references.}
The Cascade protocol, invented by Brassard and Salvail,
for instance, has the advantages of being computationally highly
efficient and also being one of the best methods in
minimizing the number of exchanged bits between Alice and Bob.
It works very well in a few percents bit error rate.
Therefore, Cascade is well suited
for implementations. Unfortunately, up till now,
a proof of unconditional security of a QKD scheme based Cascade
(and followed by, for example, standard Shor-Preskill \cite{shorpre}
or Mayers \cite{mayersqkd} privacy
amplification procedure) has been missing.
A key contribution of this paper is to provide such a proof.
The proof of security applies not only to Cascade, but to
{\it any} (interactive or non-interactive)
protocols for error correction that are based on parity computations
in QKD.

Another motivation for this work is to demonstrate
the decoupling of error correction from privacy amplification.
On the conceptual level, a QKD scheme consists of
several steps---``advantage distillation'' \cite{BBCM},
error correction and privacy amplification.
Entanglement purification has recently been proposed
by Shor and Preskill\cite{shorpre} as a useful framework for dealing
with BB84. The work of Shor and Preskill built on
earlier work in \cite{qkd} and has been subsequently
extended in \cite{twoway} to protocols involving
two-way communications and in \cite{sixstate} to the
six-state \cite{bruss} QKD scheme.

Nonetheless, an important constraint remains in those works:
The measurement operators employed by Alice and Bob must
commute {\it locally}. This local commutability constraint
ensures that those observables are simultaneous observables.
Therefore, the measurement of one observable does not
introduce any ``back-reaction'' to the measurement of
any other observables. Such a local commutability constraint
means that in analyzing QKD, one has to study both
error correction and privacy amplification together and
ensure that the observables that Alice and Bob measure {\it do}
commute locally. Therefore, this constraint complicates the
analysis.

Analysis of protocols of QKD would be greatly simplified if
one could divide up its procedure into different components
and analyze each component {\it independently}.
A main contribution of this paper is to show that
such a decoupling is, in fact, possible for error correction
and privacy amplification. The upshot is that, one
can study error correction and pick the best that
one can find. Then, one studies privacy amplification and
pick the best that one can find. Finally, one puts the two
together and the composite will remain good.
This result is reminiscient of the decoupling of source
coding from error correction in classical coding
theory.\footnote{Actually, the decoupling result in classical
coding theory is stronger than what I have stated here.
It shows that the combined protocol is {\it optimal},
even in the case of a finite block size. In contrast,
no claim of optimality for the decoupling result for QKD is
claimed in the present paper.
The issue of optimality is beyond the scope of the current paper.}

\section{BB84}
The best-known QKD scheme is BB84, in which the sender, Alice, prepares
and sends to the receiver, Bob,
a sequence of single photons randomly in one of the four
polarizations, horizontal, vertical, 45-degrees and 135-degrees.
Bob then performs a measurement randomly one of the two polarization
bases---rectilinear
and diagonal. BB84 is an example of standard ``prepare-and-measure''
protocols, which can be executed without quantum computers.
Proving the security of BB84 against the most general
attack by the eavesdropper, Eve, turned out to be a hard problem.

\subsection{entanglement purification based QKD}
Entanglement purification \cite{BDSW} has become a useful proof technique.
Consider the following entanglement purification based
QKD scheme. Alice prepares
a sequence of say $2N$ EPR pairs and sends half of each pair to Bob.
Owing to channel noises and eavesdropping attacks, those pairs will be
corrupted. Alice and Bob randomly sample say $N$ of their pairs to
estimate the
error rates in the two bases. If the error rates are too high,
they abort. Otherwise, they now apply a so-called
entanglement purification protocol
(EPP) $C$, which distills from the $N$ remaining impure
pairs a smaller number, say $m$, of almost perfectly entangled
EPR pairs. They then measure those pairs to generate a secure key.

First of all, suppose Alice and Bob share $m$ nearly perfect EPR pairs
and generate a key by measuring them. The following
theorem shows that Eve cannot have much information on the key.

\begin{thm}[\cite{qkd}]
\label{goodEPR}
If a density matrix $\rho$ has high fidelity $F$ to a state of
$m$ perfect EPR
pairs, and Alice and Bob produce their key by measuring individual
qubits of $\rho$, then with high probability, Alice and Bob have
identical $m$-bit strings $k$ with a uniform distribution, and Eve has
essentially no information about $k$. In fact, if $F \rightarrow 1$
exponentially with $m$, then Eve's information approaches $0$
exponentially with $m$ as well.\footnote{As discussed in
\cite{eff}, if we demand that Eve's information is bounded by some
small number independent of $k$, then the number of test particles
only scales logarithmically with $k$.}

\end{thm}

{\it Definition: Bell-basis}. Given a pair of qubits, a convenient
basis to use is the Bell-basis, which has Bell states as its
basis vectors. The Bell states are of the form:
\begin{equation}
\Psi^{\pm} = {  1 \over \sqrt{2} } ( | \uparrow \downarrow \rangle \pm
                                     | \downarrow \uparrow \rangle )
\end{equation}
and
\begin{equation}
\Phi^{\pm} = {  1 \over \sqrt{2} } ( | \uparrow \uparrow \rangle \pm
                                     | \downarrow \downarrow \rangle ).
\end{equation}
It is convenient to label them by two bits such that:
\begin{eqnarray}
\Phi^+ &=& 00 \nonumber \cr
\Psi^+ &=& 01 \nonumber \cr
\Phi^- &=& 10 \nonumber \cr
\Psi^- &=& 11 .
\label{bellbasis}
\end{eqnarray}

{\it Definition: N-Bell basis and BDSW notations}. Suppose Alice
and Bob share $N$
pairs of qubits. A convenient basis to use the $N$-Bell basis.
That is to say, each basis vector is the tensor product state of
$N$ Bell basis vectors. Following Eq.~\ref{bellbasis}, it is
convenient to label an $N$-basis vector by $2N$ bits. 
This is the notation employed by Bennett, DiVincenzo, Smolin and
Wootters (BDSW) \cite{BDSW}.

{\it Definition: Pauli operator}. A Pauli operator, ${\cal P}$, is defined
as a
tensor product of single-qubit
operators of the form $I$ (the identity), $X$, $Y$ and $Z$
where $X = \pmatrix{0 & 1 \cr 1 & 0}$, $Y = \pmatrix{0 & -i \cr i & \ 0}$
and $Z=\pmatrix{1 & \ 0 \cr 0 & -1}$.

{\it Definition: Stabilizer}. An Abelian group whose
generators are Pauli operators is called
a stabilizer group.

{\it Definition: Correlated Pauli strategies}. An eavesdropper,
Eve, is said to be employing a correlated Pauli strategy
if she applies a Pauli operator, ${\cal P}_i$,
to the quantum signals with some probability $p_i$.

{\it Definition: Symmetric stabilizer-based EPP}. An EPP is called
symmetric, stabilizer-based if it
involves Alice and Bob measuring operators that are
the generators of some stabilizer group.

While Eve may use any eavesdropping strategy, the following
theorem states essentially that, to consider security, one
only needs to consider correlated Pauli strategies.

\begin{thm} (Adapted from \cite{qkd})
\label{sampling}
Suppose Alice creates $M$ EPR pairs and sends
half of each to Bob.
Alice and Bob then test the error rates, $p_X$ and $p_Z$,
along the $X$ and $Z$
bases for randomly chosen disjoint subsets, $s_1$ and $s_2$, each of $m \ll M$
objects respectively.
If the error rate is too high,
they abort. Otherwise, they peform an EPP $\cal C$
on the remaining $N= M-2m$ pairs to try to distill out
$k$ EPR pairs of high fidelity.
Suppose, the EPP $C$ can correct up to $N (p_X +  \epsilon)$ phase errors
and up to $N( p_Z + \epsilon) $ bit-flip errors.
Define a Hilbert subspace ${\cal H}_{good}$ of the $N$ EPR pairs to be
the subspace spanned by $N$-Bell-states
with good error patterns. (i.e., with up to
$N (p_X +  \epsilon)$ phase errors
and up to $N( p_Z + \epsilon) $ bit-flip errors).
Let us denote the projection operator into ${\cal H}_{good}$ by $ \Pi $.
Then, we have the following:

Given {\it any} eavesdropping strategy, ${\cal S}_1$, by Eve,
there exists
a correlated Pauli strategy, ${\cal S}_2$, by Eve that will yield exactly the
same values to the following two important quantitites:

(i) $P( {\rm verification~test~is~passed~by~the~test~sample}~|~s_1, s_2) $

and 

(ii) $tr( \Pi \rho)$,

for {\it all} choices of $s_1$, and $s_2$.

\end{thm}

{\it Sketch of Proof}: The ``commuting observables''
idea in \cite{qkd} is employed. An eavesdropping strategy is
defined by the choice of an ancilla and the unitary transformation
between the combined system of the ancilla and the $N$ EPR pairs.
Given any eavesdropping strategy $S_1$ by Eve, let us
consider a fixed but arbitrary choice of sampling subsets, $s_1$, and $s_2$.
Let $O_{s_1, s_2}$ be the observable that determines whether the
verification test is passed. Recall that $\Pi$ is defined as
the projection operator into the good (i.e., correctable) Hilbert space.
Consider also $W$, the observable that gives the $2M$-bit string
respresenting the state $w$ in the BDSW notation.
Since all the observables, $O_{s_1,s_2}$'s,
$\Pi$'s and $W$ are simultaneously diagonalizable in the
$M$-Bell basis, they all commute with each other. Therefore,
it is mathematically consistent to assign probabilities to
the simultaneous eigenvalues of those observables, thus
giving rise to the two quantities
$P( {\rm verification~test~is~passed~by~the~test~sample}~|~ s_1, s_2) $
and $tr( \Pi \rho )$ for all possible choices of $s_1$, and $s_2$.

Now, imagine applying a hypothetical measurement $W$ to Alice and
Bob's state before the measurements of $O_{s_1,s_2}$'s and
$\Pi$'s. Given that $W$ commutes with 
$O_{s_1, s_2}$'s and $\Pi$'s, a prior measurement of $W$
in no way effects the outcomes of measurements of
$O_{s_1, s_2}$'s and $\Pi$'s. In other words, if Eve pre-measures
the state in the $N$-Bell-basis (i.e., measures $W$),
neither
the probability of passing the verification test, nor the
probability of being in the good Hilbert space will be affected
by such a prior measurement. However, with such a prior measurement,
Eve has reduced her eavesdropping strategy $S_1$ to a correlated
Pauli's strategy, $S_2$.

{\it Remark}: This commuting observables idea applies to all symmetric
stabilizer-based EPPs
including ones that involve two-way classical communications.

Theorem~\ref{sampling} is telling us that one can treat
the two important quantities---i) the probability of passing
a verification test and ii) the probability of being in a good Hilbert
space, $tr( \Pi \rho)$---as {\it classical}.
In essence, one can apply classical sampling theory to
a quantum problem. Furthermore,
$tr( \Pi \rho)$ provides a
bound to the fidelity of the corrected EPR pairs:

\begin{thm}[\cite{shorpre,squeeze}]
\label{goodhilbertspace}
Consider a stabilizer-based EPP $\cal C$ which distills
$m$ EPR pairs from $n$ impure pairs.
Suppose $\cal C$ works perfectly in a Hilbert subspace
${\cal H}_{good}$, which is
spanned by Bell-states with good error patterns (i.e., correctable
by $\cal C$). Denote the projection operator onto ${\cal H}_{good}$
by $\Pi$. If we apply the EPP $\cal C$ to an initial state
$\rho$, then the fidelity of the recovered state as $m$ EPR pairs
is bounded below by
\begin{equation}
F \equiv \langle {\bar{\Phi}}^{(m)}
| \rho_{rec.} | \bar{\Phi}^{(m)} \rangle \geq tr( \Pi \rho) .
\end{equation}

Here, $\rho_{rec.}$ is the recovered state after error correction,
$ {\bar{\Phi}}^{(m)}$ is the $m$-EPR pair state.
\end{thm}

{\it Proof}: 
This Theorem follows from standard stabilizer
quantum error correcting code (QECC) theory. An explicit proof
of essentially the same result can be found in \cite{squeeze}.~Q.E.D.

\subsection{reduction to BB84 via CSS codes}

Because of Theorem~\ref{goodhilbertspace}, EPP
based QKD schemes are particularly convenient to analyze.
Unfortunately, they are difficult to implement because
they generally require Alice and Bob to possess quantum
computers. A key insight of Shor and Preskill is to remove
the requirement of quantum computers by showing
that, in fact, the security of a special class of EPP based QKD
schemes implies the security of BB84. More concretely,
they considered a special class of
quantum error-correcting codes, called Calderbank-Shor-Steane
(CSS)\cite{CS,steane}
codes (see below for properties of CSS codes)
and proved the following theorem:

\begin{thm}[\cite{shorpre}]
\label{thm:reduction}
Given an EPP-based QKD scheme that is based on a
CSS code and a verification procedure that involves only two
bases, its security implies the security of a BB84 scheme.
\end{thm}

{\it Remark}: Similarly, when the verification procedure involves
three bases, an analogous Theorem shows that the security of an EPP-based
QKD scheme that is based on a CSS code implies
the security of the six-state scheme.

We shall refer the readers to \cite{shorpre,squeeze} for details of
the proof of Theorem~\ref{thm:reduction}.
A CSS code is a stabilizer-based quantum code with generators
that are either i) tensor products of the identities and $Z$'s only
or ii) tensor products of the identities and $X$'s only. It has the
advantage that the phase and bit-flip error correction procedures
are totally decoupled from each other.\footnote{Applying
an operator $X$ to a state will introduce a bit-flip error
to the state. Similarly, applying an operator $Z$ to a state will
introduce a phase error. Finally, applying an operator $Y$ will
lead to both a bit-flip and a phase error.

The intuitive reason
why an EPP-based QKD can be reduced to BB84 is that Alice and
Bob do not need to compute or announce their phase error syndromes.
This is because the phase errors do not affect the value of
the final key. Roughly speaking, randomizing the state over all possible
phase error syndromes, one recovers BB84.
In other words, provided that, from Eve's point of view, Alice and Bob
{\it could have} performed the QKD scheme by quantum computers,
the resulting BB84 scheme is secure. Alice and Bob do not
really have to use quantum computers.}

More concretely, a CSS code is defined as follows:
Consider a binary linear classical code $C_1$ and its subcode $C_2$.
A codeword of a CSS code is an equal superposition of codewords of
$C_1$ that are in the same coset of $C_2$:

\begin{equation}
| \phi_u \rangle = \sum_{ v \in C_2 } | u + v \rangle.
\end{equation}
Note that, if $u_1 - u_2 \in C_2$, then $ | \phi_{u_1} \rangle =
| \phi_{u_2} \rangle$. Therefore, the codeword of a CSS code is
in one-one correspondence with the cosets of $C_2$ in $C_1$.
Suppose both $C_1$ and the dual of $C_2$, $C^{\perp}_2$, can
correct up to $t$ errors. Then, the CSS code based on $C_1$ and
$C_2$ can correct up to $t$ bit-flip errors and $t$ phase errors.

On reduction from EPP to BB84, the EPP leaves its mark as
an error correction/privacy amplification protocol in the
following manner. Alice sends a random quantum state
$ | w \rangle$ to Bob. Owing to noises in the channel and
eavesdropping actions, Bob receives it as a corrupted string $ w + e$.
Afterwards, Alice picks a random
codeword $u \in C_1$ and broadcasts $ w + u$. Bob substracts
this from his string to obtain $u + e$. He then corrects
the error to obtain $u$. Finally, he generates the key as the
coset $u + C_2$. Notice that, the cosets of a code, say $C_2$,
is in one-one correspondence with the error syndromes. Indeed,
the value of the key is given by the error syndrome of the subcode
$C_2$ for a codeword in $C_1$.

Using CSS codes and Theorem~\ref{thm:reduction},
BB84 is proven to be secure up to an error rate of
$11$ percents. By using two-way classical communications,
BB84 can be made
secure at a much higher error rate of about $17$
percents.\footnote{Note that it has been shown that BB84 with
only one-way classical communications is necessarily insecure
at an error rate of about $15 \%$ \cite{fuchs,cirac}.
Therefore,
this result in \cite{twoway} shows
clearly that BB84 with two-way classical communications is definitely
better than BB84 with only one-way classical communications.}
This is due to
the following theorem by Gottesman and myself \cite{twoway}, 
hich generalizes
Theorem~\ref{thm:reduction}.

\begin{thm}[\cite{twoway}]
\label{main}
Suppose a two-way EPP satisfies the
following conditions:

\begin{enumerate}

\item \label{symmetric} (Symmetric) It can be described as a
series of measurements
$M_i$, with both Alice and Bob measuring the same $M_i$.

\item \label{CSSlike} (CSS-like) Each of its generators $M_i$ can be
written as
either a) a product of $X$'s only or b) a product of $Z$'s only.

\item \label{locallycommuting} (Locally-commuting) Each pair of
$M_i$ and $M_j$ commute locally in Alice's (or Bob's) side.

\item \label{conditional} (Conditional on $Z$'s only)
All conditional operations depend on the result of measuring
$Z$ operators only.

\end{enumerate}

Let us call such a protocol a {\it reducible} protocol.
Claim: a reducible protocol {\it can} be converted to a standard
``prepare-and-measure'' QKD scheme with security equal to the
EPP-based QKD scheme.
\end{thm}

{\it Remark}: Here, the notation has been slightly abused.
By a products of $Z$'s only, I actually mean a product of
the identities and the $Z$'s only. Similarly, for $X$'s.

{\it Remark}: If the verification stage involves two bases,
then the ``prepare-and-measure'' QKD scheme is BB84.
If it involves three bases, then the ``prepare-and-measure''
QKD scheme is the six-state scheme.

We will refer the readers to \cite{twoway} for the details of the proof
of Theorem~\ref{main}.
\section{Constraint on local commutability}

Theorem~\ref{main} is a strong result in QKD. Nonetheless,
the constraint \ref{locallycommuting} in Theorem~\ref{main} seriously
restricts its applicability. In the EPP picture, the constraint
demands that all the local measurement operators
that Alice and Bob employ must {\it commute locally}
with each other. Therefore, one is not at liberty to
choose the bit-flip and phase error correction measurement
operators independently.

I remark that the local commutability constraint is
a big obstacle in the application of Theorem~\ref{main} to prove the
security of interactive Cascade scheme \cite{cascade} for error correction
proposed by Brassard and Salvail. Recall the Cascade protocol
involves a binary search subroutine, ``BINARY'', by Alice and Bob,
which allows them to identify the location of an error.
The binary search subroutine, BINARY,
involves the computation of the parity of a set
and subsequently dividing it into two sets and
computing the parity of each subset, etc, until the
location of the error is found.
Note that at the end of BINARY, the size of a subset is
reduced to a single object, which means Alice (and also Bob) has to
announce the eigenvalue $Z_i$ of a {\it single} qubit at
location $i$ (i.e., the $i$-th qubit). Now,
any quantum error correcting procedure that corrects the
phase error of the announced bit must contain a measurement operator $M$
with a component $X_i$ for also the $i$-th qubit.
This means that $M$ anti-commutes, rather than commutes
with $Z_i$. In conclusion, with Cascade protocol, it
would be impossible to correct all the phase errors.
Therefore, the application of Theorem~\ref{main} to the
Cascade protocol looks problematic.

\subsection{Using ancillary EPR pairs}
To resolve this problem of local non-commutability, notice
that a) all symmetric measurement operators, $M_i = M_i^A \otimes M_i^B$
{\it do} commute globally and b) in many cases, only this {\it relative}
error syndrome between Alice and Bob is of interest.
For instance, in BINARY, Alice and Bob are interested in
only whether their corresponding parities agree or disagree, but not in
the actual values of the individual parities.
A simple method to bring two
distant quantum systems together and allow a {\it global} operator to
be measured is teleportation. To achieve teleportation, some
ancillary EPR pairs must be shared by Alice and Bob.
This motivates the basic insight of the current paper---to use
ancillary EPR pairs to compute the relative
error syndrome.

Instead of teleportation, a more efficient way of measuring the global
error syndrome will be employed. Here is a main theorem of the
current paper.

\begin{thm}
\label{main2}
Suppose Alice and Bob share a number of impure EPR pairs
and they would like to
compute $r$ symmetric global operators each
of the form $M_i = M^A_i \otimes M^B_i$
(As before, by symmetric, it means that
$M^A_i$ is the same as $M^B_i$ except that they act on
Alice's and Bob's Hilbert spaces respectively) and $M^A_i$ is a Pauli
operator. Suppose further that they would like to know only the
eigenvalues of $M_i$'s, but otherwise leave the state
unchanged. The claim is that they can do so with $r$ {\it ancillary}
EPR pairs.

\end{thm}

{\it Sketch of Proof}: The notation is such that an EPR
pair is an eigenstate of $ZZ$ and $XX$, with eigenvalue $+1$ for both.
Let us call the two qubits of the $j$-th EPR pair shared by Alice and Bob,
$A'_j$ and $B'_j$ respectively. For each operator, $M_i$,
Alice measures $M^A_i \otimes Z_{A'_i}$ and broadcasts her
outcome and Bob measures $M^B_i \otimes Z_{B'_i}$ and broadcasts
his outcome. The relative outcome, the product of
$ M^A_i \otimes Z_{A'_i} \otimes  M^B_i \otimes Z_{B'_i}$ gives
the eigenvalue of the operator $M_i$ (because the state of the ancillary
EPR pair gives an eigenvalue $+1$ for the operator
$ Z_{A'_i} \otimes Z_{B'_i}$). More importantly,
by an explicit calculation 
analogous to the argument in teleportation, one can
show that no disturbance to the state is made except for the
determination of the eigenvalue of $M_i =  M^A_i \otimes M^B_i$.~Q.E.D.

The above theorem employs a generalization of the so-called breeding
method for EPP, studied in \cite{BBPSSW} (see also \cite{BDSW}).
In \cite{BDSW}, the breeding method was only mentioned on
passing because it had been superseded by the standard
hashing method, which can be performed without ancillary
EPR pairs. Let me call a general EPP that involves ancillary
EPR pairs a {\it generalized} breeding protocol/method.
In contrast to prior art, here I notice that the generalized
breeding protocol is, in general, {\it not} reducible to
a non-breeding protocol. In fact, it is more powerful because it
allows the decoupling of
error correction from privacy amplification.
In summary, the decoupling of error correction from privacy amplification
is achieved at the price of introducing ancillary EPR pairs
shared by Alice and Bob.

I remark that the calculation of $M^A_i \otimes Z_{A'_i}$
(and similarly $M^B_i \otimes Z_{B'_i}$)
in Theorem~\ref{main2} can, indeed, be done
by local quantum gates. The actual quantum circuit diagram is
very similar to the ones discussed in
for example, \cite{BDSW} and \cite{BBPSSW}.
Since the actual construction is outside the main theme of this paper,
the details will be skipped here.

\subsection{Reduction to BB84}

Using ancillary EPR pairs in a generalized breeding protocol, the
above subsection shows that one can decouple error correction from
privacy amplification in a QKD scheme. However, such a scheme
generally requires a quantum computer to implement. So, the
next question is: how to reduce the above protocol to standard BB84?
Here is the second main Theorem of the current paper.

\begin{thm}
\label{main3}

Suppose a two-way EPP satisfies the
following conditions:

\begin{enumerate}

\item (Symmetric) It can be described as a
series of measurements
$M_i$, with both Alice and Bob measuring the same $M_i$.

\item (CSS-like) Each of its generators $M_i$ can be
written as
either a) a product of $X$'s only or b) a product of $Z$'s only.
Let me call them $M_X$ and $M_Z$ operators respectively.

\item \label{rlocallycommuting} (r-locally-non-commuting)
There exists a set of $r$ $M_Z$ operators, which, thereafter
I shall call the {\it non-commuting
set} such that, after deleting them
from the set of measurements, each pair of operators
$M_j$ and $M_k$ chosen from the remaining set of measurements
commute locally in Alice's (or Bob's) side.

\item (Conditional on $Z$'s only)
All conditional operations depend only on the result of measuring
$Z$ operators.

\end{enumerate}

Then the protocol {\it can} be converted to a standard
``prepare-and-measure'' QKD scheme with security equal to the
EPP-based QKD scheme, {\it provided that} Alice and Bob initially
share an $r$-bit secret string and use it to encode the
measurement outcome of $M_Z$'s of the non-commuting set in
Condition~\ref{rlocallycommuting}.\footnote{Note that the
same key is used to encode the measurement outcomes in both Alice
and Bob's sides. This is because the relative error syndrome is
{\it allowed} to be disclosed to Eve.}\footnote{Note that the final key
is now a coset of $C_2$ in $F^n_2$, whereas in Shor-Preskill's proof,
the key is a coset of $C_2$ in $C_1$. The difference is due to the
fact that, in Theorem~\ref{r-reducible}, an ancillary secret is sacrified.
The {\it net} key generation rate is the same if Theorem~\ref{r-reducible}
is applied in lieu of Shor-Preskill's proof.}
\label{r-reducible}

\end{thm}

{\it Sketch of Proof}: Combine the proofs of Theorems~\ref{main}
and \ref{main2}. In other words, the proof of Theorem~\ref{main2}
can be used to relax the constraint of local commutability
in Theorem~\ref{main}, thus giving Theorem~\ref{main3}.

We have the following Corollary:

\begin{cor}
Consider the purification of $N$ impure EPR pairs. Suppose one is given
a symmetric stabilizer-based bit-flip (interactive or one-way)
error correction procedure with $s$ operators $M_Z$'s and also
a symmetric stabilizer-based phase error correction procedure
with $t$ operators $M_X$'s acting on
the $N$ pairs.

Claim: The combined error correction/privacy amplification protocol can
be reduced to a standard prepare-and-measure QKD protocol, provided
that Alice and Bob initially share an $s$-bit secret string.
Having sacrificed the initial $s$-bit string,
the output of the procedure is an $N-t $-bit secret string.\footnote{See
footnote~9.}
\label{r-corollary}
\end{cor}

{\it Remark}: As an application of the above Corollary,
the following protocol for error correction/privacy amplification
of QKD is unconditional secure:
Step~1: the Cascade scheme for error correction, modified by the one-time-pad
encryption of its bit-flip error syndrome, followed by
Step~2: a random hashing
procedure \cite{mayersqkd,shorpre}. Notice that this
is a rather efficient protocol in terms of both the key generation rate and
computational power.

For schemes involving concatenation, there is the following Corollary:

\begin{cor}
Suppose an EPP, $\cal C$ is a concatenation of two subroutines,
$S_1$ and $S_2$, where the first subroutine, $S_1$ satisfies all
the conditions in Theorem~\ref{main} (i.e., symmetric, CSS-like,
locally-commuting and conditional on $Z$'s only)
and the second subroutine, $S_2$ satisfies Theorem~\ref{r-reducible}
as an $r$-locally-noncommuting (symmetric, CSS-like, conditional on $Z$'s
only) EPP.
Then, the protocol $\cal C$ can be converted to a prepare-and-measure
QKD protocol
with the same security, provided that Alice and Bob initially share
an $r$-bit secret string and use it for one-time-pad encryption of
the measurement outcomes of the $r$ pairs\footnote{See footnote~8.}
of measurement outcomes in the
non-commuting set.

\label{r+1-corollary}
\end{cor}

The upshot of the above Corollary is that the decoupling result
remains valid even when there are two way classical
communications\cite{twoway} and even when concatenated codes are
employed.

\section{Concluding Remarks}
In summary, I have considered a rather general class of entanglement
purification schemes, more specifically,
symmetric, stabilizer-based schemes and their reduction to
BB84. It was shown that in those
schemes, the procedure for error correction can be decoupled from
the procedure for privacy amplification. The decoupling is achieved
by requiring Alice and Bob to share a modest initial string
and use it for the one-time-pad encryption of the bit-flip
error syndrome. This is no change in the net key generation rate
because the loss of this initial string will
be exactly compensated by the generation of a longer key. (See
footnotes~8 and~9.)
As a corollary, I prove the security of the Cascade scheme,
modified by one-time-pad encryption of error syndrome,
followed by a random hashing privacy amplification procedure.
This is an efficient scheme in terms of both key generation
rate and computational power.

\section{acknowledgement}
I particularly thank Norbert L\"{u}tkenhaus for bringing to
my attention the question of the
security proof of the Cascade scheme and for many enlightening discussions.
Helpful conversations with colleagues including Daniel Gottesman,
Tsz-Mei Ko,
John Preskill and Peter Shor
are also gratefully acknowledged.


\end{document}